\documentclass[prl,aps,twocolumn,a4paper,nofootinbib]{revtex4-1}

\newif\ifusesec
\usesectrue 
   
\usepackage{graphicx,psfrag}
\usepackage{mathrsfs}
\usepackage{amsmath,amsfonts,amssymb}
\usepackage{multirow}
\usepackage{comment}
\usepackage{hyperref}
\hypersetup{
  colorlinks=true,        
  linkcolor=blue,         
  citecolor=red,          
  urlcolor=red,          
}
\DeclareSymbolFontAlphabet{\mathrsfs}{rsfs}
\DeclareMathAlphabet\mathbfcal{OMS}{cmsy}{b}{n}

\newcommand{\be}{\begin{equation}}
\newcommand{\ee}{\end{equation}}
\newcommand{\bea}{\begin{eqnarray}}
\newcommand{\eea}{\end{eqnarray}}
\newcommand{\bel}{\begin{align}}
\newcommand{\eel}{\end{align}}

\def\DErad{{\Delta E^{\rm NR}}}
\def\DJrad{{\Delta J^{\rm NR}}}

\def\GMc2{G M_{\odot} c^{-2}}

\def\F{{\cal F}}

\def\F{{\cal F}}

\DeclareSymbolFontAlphabet{\mathrsfs}{rsfs}
\DeclareMathAlphabet{\mathcal}{OMS}{cmsy}{m}{n}

\usepackage{color}

\definecolor{cyan}{rgb}{0,0.9,0.9}
\definecolor{orange}{rgb}{0.9,0.5,0}
\definecolor{magenta}{rgb}{1,0,1}
\definecolor{purple}{rgb}{0.8,0.4,0.8}
\definecolor{gray}{rgb}{0.8242,0.8242,0.8242}

\begin{document}

\title{Strong-Field Scattering of Two Black Holes: Numerics Versus Analytics}

\author{Thibault \surname{Damour}$^1$}
\author{Federico  \surname{Guercilena}$^{2,3}$}
\author{Ian  \surname{Hinder}$^2$}
\author{Seth  \surname{Hopper}$^2$}
\author{Alessandro \surname{Nagar}$^1$}
\author{Luciano  \surname{Rezzolla}$^{3,2}$}
\affiliation{$^1$Institut des Hautes Etudes Scientifiques, 91440 Bures-sur-Yvette, France}
\affiliation{$^2$Max-Planck-Institut f{\"u}r Gravitationsphysik,
         Albert-Einstein-Institut, Am M\"uhlenberg 1,
         D-14476 Golm, Germany}
\affiliation{$^3$Institut f\"ur Theoretische Physik, 
         Max-von-Laue-Str. 1, D-60438 Frankfurt am Main, Germany}

\begin{abstract}
We probe the gravitational interaction of two black holes in the
strong-field regime by computing the scattering angle $\chi$ of
hyperbolic-like, close binary-black-hole encounters as a function of the
impact parameter. The fully general-relativistic result from numerical
relativity is compared to two analytic approximations: post-Newtonian theory
and the effective-one-body formalism. 
As the impact parameter decreases, so that black holes pass within a 
few times their Schwarzschild radii, we find that the post-Newtonian prediction
becomes quite inaccurate, while the effective-one-body one keeps showing
a good agreement with numerical results. 
Because we have explored a regime which is very different from the one 
considered so far with binaries in quasi-circular orbits, our results 
open a new avenue to improve analytic representations of the 
general-relativistic two-body Hamiltonian.
\end{abstract}

\date{\today}

\maketitle

\paragraph*{Introduction--}

Historically, elastic scattering experiments have been an essential tool
to explore fundamental interactions in nature. Following this tradition,
we have performed systematic numerical experiments exploring the
gravitational interaction of two black holes (BHs) in a regime where
strong-field effects become important. In this {\it Letter} we report on
the first numerical-relativity (NR) computation of the gauge-invariant
dynamical relation between the (center of mass) {\it scattering angle}
$\chi$ of close binary-black-hole (BBH) encounters and the energy $E$ and
angular momentum $J$ of the system. Differently from what has been done
in previous
works~\cite{Pretorius:2007jn,Shibata:2008rq,Sperhake:2008ga,Sperhake:2009jz,Witek:2010xi,Sperhake:2012me},
which have looked at the {\it strongly inelastic} collision of two BHs
leading to immediate (or prompt) merger, we here concentrate on what is
an essentially (hyperbolic-like) {\it elastic} scattering.

More specifically, we study a set of configurations in which two
equal-mass, nonspinning BHs start at large separations
$\sim 100\,GM/c^2$ 
with mildly relativistic individual velocities
$|v_1|/c=|v_2|/c\approx 0.21$,
approach each other within a few times their Schwarzschild radii, and
then separate again towards infinity. By varying the initial angular
momentum (or, equivalently, the impact parameter) we explore a large
range of scattering angles from 70.7 degrees up to 305.8 degrees. We then
compare the value of $\chi$ determined from NR to several analytical
estimates: post-Newtonian (PN) theory and the effective-one-body (EOB)
formalism~\cite{Buonanno:1998gg,Buonanno:2000ef,Damour:2000we}, finding
that PN becomes inaccurate for small values of the impact parameter, while EOB
continues to show good agreement.


\begin{figure}[h]
  \centering
  \includegraphics[width=.5\textwidth]{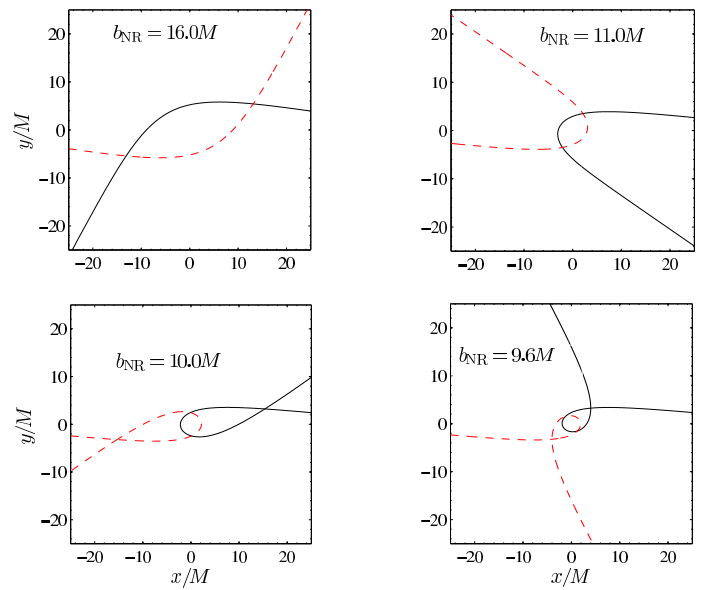} 
    \caption{\label{fig:tracks} Coordinate trajectories of the two BHs in
      hyperbolic-like encounters for four selected values of the impact
      parameter ${b}_{\rm NR}$.}
\end{figure}

\begin{table}[t]
  \caption{\label{tab:all_info} From left to right, the columns report:
    the NR ``impact parameter'', the initial energy, the initial 
    angular momentum, the gravitational-wave energy and
    angular-momentum losses.}
  \begin{center}
    \begin{ruledtabular}
\begin{tabular}[c]{lllll}
 ${b}_{\rm NR}/M$ &$~~E^{\rm NR}_{\rm in}/M$ & $~~J^{\rm NR}_{\rm in}/M^2$  &  $\Delta{E}^{\rm NR}/M$  & $\Delta{J}^{\rm NR}/M^2$  \\
\hline
9.6  & \!\!\!\!\!\!\!1.0225555(50) & 1.099652(36) & 0.01946(17)    & 0.17007(89)\\
9.8  & \!\!\!\!\!\!\!1.0225722(50) & 1.122598(37) & 0.01407(10)    & 0.1380(14) \\
10.0 & \!\!\!\!\!\!\!1.0225791(50) & 1.145523(38) & 0.010734(75)   & 0.1164(14) \\
10.6 & \!\!\!\!\!\!\!1.0225870(50) & 1.214273(40) & 0.005644(38)   & 0.076920(80)\\
11.0 & \!\!\!\!\!\!\!1.0225884(50) & 1.260098(41) & 0.003995(27)   & 0.06163(53) \\
12.0 & \!\!\!\!\!\!\!1.0225907(50) & 1.374658(45) & 0.001980(13)   & 0.04022(53) \\
13.0 & \!\!\!\!\!\!\!1.0225924(50) & 1.489217(48) & 0.0011337(90)  & 0.029533(53)\\
14.0 & \!\!\!\!\!\!\!1.0225931(50) & 1.603774(52) & 0.0007108(77)  & 0.02325(47) \\
15.0 & \!\!\!\!\!\!\!1.0225938(50) & 1.718331(55) & 0.0004753(75)  & 0.01914(76) \\
16.0 & \!\!\!\!\!\!\!1.0225932(50) & 1.832883(58) & 0.0003338(77)  & 0.0162(11)  \\
\end{tabular}
  \end{ruledtabular}
\end{center}
\end{table}
\begin{table*}[t]
  \caption{\label{tab:EOBNRPN}NR, EOB and PN estimates of the scattering
    angle $\chi$ at different PN orders. Angles are measured in degrees.}
  \begin{center}
    \begin{ruledtabular}
      \begin{tabular}{cllccccclll}
 ${b}_{\rm NR}/M$  &  $\hat{r}_{\rm min}^{\rm EOB}$ & $\chi^{\rm NR}$ & $\chi^{\rm EOBNR}_{\rm 5PNlog}$ & $\chi^{\rm EOB}_{\rm 4PN}$  &   $\chi^{\rm EOB}_{\rm 3PN}$ & $\chi^{\rm EOB}_{\rm 2PN}$  &  $\chi^{\rm EOB}_{\rm 1PN}$ &
$\chi^{\rm PN}_{\rm 3PN}$ & $\chi^{\rm PN}_{\rm 2PN}$ & $\chi^{\rm PN}_{\rm 1PN}$ \\
\hline
9.6  & 3.3   & 305.8(2.6) & 322(62)     & 364.29   &   $\dots$ &   $\dots$   & $\dots $     &  139.9     & 124.2    & $\dots$ \\
9.8  & 3.7   & 253.0(1.4) & 261(14)     & 274.92   &    332.24 &   $\dots$   & $\dots $     &  131(2)    &  118.46  & $\dots$ \\
10.0 & 4.0   & 222.9(1.7) &   227(5)    & 234.26   &    259.46 &   $\dots$   & $\dots $     &  126(1)    &  115.89  & $\dots$ \\ 
10.6 & 4.8    & 172.0(1.4) & 172.8(7)   & 174.98   &    182.09 &   220.11    &  260.53      &  118.5(3)  &  112.43  & $\dots$ \\ 
11.0 & 5.3    & 152.0(1.3) & 152.4(3)   & 153.59   &    157.68 &   177.60    &  194.90      &  114.7(2)  &  110.14  & $\dots$ \\
12.0 & 6.5    & 120.7(1.5) & 120.77(6)  & 121.17   &    122.63 &   129.98    &  136.42      &  104.34(4) &  102.06  & $\dots$ \\
13.0 & 7.6    & 101.6(1.7) & 101.63(2)  & 101.80   &    102.48 &   106.20    &  109.80      &  93.69(2)  &    92.54 & $\dots$ \\
14.0 & 8.6    & 88.3(1.8)  &  88.348(8) & 88.43    &     88.80 &    90.95    &   93.30      & 84.111(7)  &    83.55 & $\dots$ \\
15.0 & 9.7    & 78.4(1.8)  &  78.427(4) & 78.47    &     78.69 &    80.03    &   81.699     &  75.962(3) &    75.71 & 169.298 \\
16.0 & 10.8   & 70.7(1.9)  &  70.666(2) & 70.69    &     70.84 &    71.71    &   72.951     &  69.122(2) &    69.03 & 108.894 \\
   \end{tabular}
  \end{ruledtabular}
\end{center}
\end{table*}

\paragraph*{Numerical-relativity simulations--}
\label{numerics}

The simulations were performed using the open-source Einstein
Toolkit~\cite{Loffler:2011ay} within the Cactus~\cite{Cactuscode:web}
framework, using the McLachlan~\cite{Brown:2008sb} evolution code and
8th-order spatial finite differencing.
%
%
The computational domain extends to $400\,M$ in units in which $G=c=1$, and
outgoing radiative boundary conditions are used at the outer boundary. Here $M = m_1+m_2$ where
$m_1=m_2$ is the mass of each BH, as determined from the apparent
horizons using the AHFinderDirect~\cite{Thornburg:2003sf} code. 
The domain is discretised with a Cartesian numerical grid and 7 
levels of box-in-box grid refinement around each BH provided by the adaptive-mesh-refinement code
Carpet~\cite{Schnetter:2003rb}.  The refined regions track the BHs and have a finest grid spacing
of $h = 0.025\,M$ (low resolution) and $h = 0.017\,M$ (high resolution).


Initial data of the Bowen-York form is constructed using the
TwoPunctures~\cite{Ansorg:2004ds} code. The BHs start on the $x-$axis
with initial positions designated by $\pm X$ and initial momenta
$(p_x,p_y,p_z)= \pm|\vec p| (-\sqrt{1-(b_{\rm NR}/(2X))^2},b_{\rm
  NR}/(2X),0)$. Here $b_{\rm NR}$ is the NR ``impact
parameter'', which is related to the ADM angular momentum via
$J_\mathrm{\rm ADM} = 2 X |p_y| \,=\, |\vec p|\,b_{\rm NR}$. For all the
simulations we use $|\vec p| = 0.11456439\, M$, $X = 50\, M$; 
more information on the 10 initial
configurations is collected in Table~\ref{tab:all_info}. Since these
configurations vary only in the direction of the initial momentum, the
ADM energy, $E_\mathrm{\rm ADM}$, of each spacetime is nearly the same
($E_\mathrm{ADM} - M \approx 2.26\times10^{-2}M$), while the angular
momentum $J_\mathrm{\rm ADM} =\, |\vec p|\, b_{\rm NR}$ is proportional to
$b_{\rm NR}$. For our comparisons we are actually interested in the
``initial'' energy and angular momentum left after the burst of
spurious radiation present in the initial data. These quantities, that we
indicate as $(E_{\rm in}^{\rm NR},J_{\rm in}^{\rm NR})$, differ fractionally
 by only $10^{-5}$ from $(E_{\rm ADM},J_{\rm ADM})$ and are listed
in Table~\ref{tab:all_info}.

To ease the analytic computations of the scattering angle we have also
measured the total energy, $\DErad$, and angular momentum, $\DJrad$,
radiated in gravitational waves during the scattering event. We obtain
these quantities by first computing the multipolar modes (up to $\ell =
8$) of the Weyl scalar $\Psi_4$ at several finite radii. At each radius
we perform time-domain integrations of these moments and sum them to
obtain $\DErad$ and $\DJrad$.
The resulting finite-radius values of $\DErad$ and $\DJrad$ are then
extrapolated to null infinity.
We (over-)estimate the extrapolation error in these
quantities as the difference in the extrapolated value and the value at
the largest radius.

We track the motion of the BHs using the Cartesian coordinate positions of the
punctures which we convert to polar coordinates
$(r, \varphi)$. Treating the incoming $\varphi_\mathrm{in}(r)$ and
outgoing $\varphi_\mathrm{out}(r)$ paths separately, we extrapolate
$\varphi_\mathrm{in,out}(r)$ as $r \to \infty$ by fitting each of them to a
polynomial of order $n$ in $1/r$ to measure the two asymptotic angles
$\varphi^{\infty}_\mathrm{in,out}$ corresponding to a binary with
infinite separation. The total scattering angle is then calculated as
$\chi^\mathrm{NR} \equiv \varphi^{\infty}_\mathrm{out}
-\varphi^{\infty}_\mathrm{in}-\pi$. A range of $(r, \varphi)$ must 
be chosen to perform the polynomial fitting and compute 
$\varphi^{\infty}_\mathrm{in,out}$. For
the incoming and outgoing paths
we extrapolate from $r \in [18.75,75]\,M$ and $[25,100]\,M$, respectively.
With this choice, we extrapolate over a range in $1/r$ that is 1/4 the
size of the data we use for the fitting. Our least-squares fitting method
employs a singular-value decomposition (SVD) which drops singular values
smaller than a threshold (chosen to be $10^{-13}$ times the
maximum singular value). We then choose the polynomial order $n$ as the
largest for which the SVD threshold allows variations in the constant
term. 
We take as our estimate of the extrapolation error the maximum difference 
between the extrapolant at order $n$ and the extrapolant at all orders 
between 1 and $n-1$. We have tested that this error estimate is robust
with respect to variations in the details of the extrapolation method.
We expect that the extrapolated scattering angle will be
insensitive to the details of the 
spatial gauge conditions
employed due to their symmetry-seeking nature and the fact that the
region between the far-separated BHs is approximately Minkowskian.

We estimate our finite-difference error in $\chi$, $\DErad$, and $\DJrad$
by performing each simulation with two different resolutions and
(conservatively)  assuming
4th-order convergence.
The total error estimates, as shown in Table~\ref{tab:all_info}, are computed
by adding the finite-difference and extrapolation errors in quadrature.
Note that we also explored other sources of error (e.g., the effect of finite
initial separation and the choice of how much data to use in the
fitting), but found that these were all much smaller.




\begin{figure}[t]
  \centering
 \includegraphics[width=.47\textwidth]{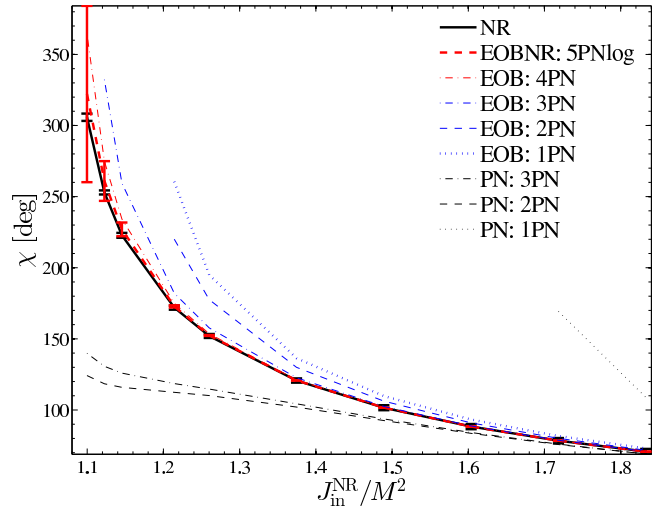} 
    \caption{\label{fig:angle} Comparing the NR scattering angle with
      various EOB and PN predictions. NR data and the state-of-the-art EOB 
      model agree within their respective error bars.}
\end{figure}

\paragraph*{EOB and PN analytic computations of $\chi$ and comparison with NR results--}
\label{sec:analytics}

From the analytical relativity (AR) point of view, the scattering angle depends on the
full equations of motion, including both conservative (Hamiltonian $H$)
and radiation-reaction ($\F_{\rm rad\;reac}$) effects.
The current AR knowledge of $\F_{\rm rad\;reac}$ along general (non-quasi-circular)
motions~\cite{Bini:2012ji} is less complete than that of $H$ and cannot be used
for accurate NR/AR comparisons. However, it has been recently pointed 
out~\cite{Bini:2012ji} that, when neglecting terms quadratic in
 $\F_{\rm rad\;reac}$
(i.e., of order $(v/c)^{10}$, where $v$ is the velocity), the scattering
angle $\chi$ can be analytically computed solely from the knowledge of
the Hamiltonian $H$.
More precisely, the AR approximation
$\chi^{\rm AR}$ is given by the value it would have in a
conservative-dynamics scattering of a binary system whose energy and
angular momentum are the {\it average} values between the incoming and
outgoing states: 
\be
\label{eq:chi_angle}
\chi^{\rm AR}= \chi^{(\rm conservative)}(\bar{E},\bar{J}),
\ee
where $\bar{E}\equiv \left(E_{\rm in}+E_{\rm out}\right) / 2$ and
$\bar{J}\equiv \left(J_{\rm in}+J_{\rm out}\right) / 2$. Using the NR
measures of the radiative NR losses $\DErad=E_{\rm in}^{\rm NR}-E_{\rm
  out}^{\rm NR}$, $\DJrad=J_{\rm in}^{\rm NR}-J_{\rm out}^{\rm NR}$, we
have $\bar{E}=E_{\rm in}^{\rm NR}- \DErad / 2$ and $\bar{J}=J_{\rm
  in}^{\rm NR}- \DJrad / 2$.

We compute $\chi^{\rm AR}$ using various  EOB and 
PN Hamiltonians.
In all cases, we numerically integrate the equations of motion from
an initial separation $r_0=10000\,M$ up to a comparable final separation, 
and compute $\chi^{\rm AR}\equiv \varphi_{\rm final}-\varphi_{0}-\pi$.
In the following we denote $\mu\equiv m_1 m_2/M$, $\nu=\mu/M$,
$j\equiv p_\varphi\equiv J/(M\mu)$, and
$u\equiv M/r$.

The EOB conservative binary dynamics is completely encoded in 
two functions $A(u;\nu)$ and $B(u;\nu)$. The radial interaction
potential $A$ is a $\nu$-deformed generalization of the Schwarzschild
potential $A_{\rm Schw}\equiv 1-2M/r=1-2u$. 
The potentials $A$ and $B$ feed into the EOB Hamiltonian $H_{\rm
    EOB}(r,p_\varphi,p_r)\equiv M\sqrt{1+2\nu\left(H_{\rm
      eff}/\mu-1\right)}$, where $H_{\rm eff}= \mu \sqrt{A(1+j^2 u^2 +
    2\nu(4-3\nu) u^2 p_{r_*}^4) + p_{r_*}^2}$. Here $p_{r_*}\equiv
  \sqrt{A/B}\,p_r$ is a tortoise version of the $\mu$-rescaled radial
  momentum $p_r\equiv P_r/\mu$.

Currently, the analytically most complete (5PN with logs),
  NR-calibrated version of $A$ \cite{Damour:2012ky} is defined as the
Pad\'e approximant $P^1_5\left(A_{\rm 5PNlog}^{\rm Taylor}\right)$ with
$A_{\rm 5PNlog}^{\rm Taylor}\equiv 1-2u+2\nu u^2 +\nu a_4 u^3 + \nu
(a_5^c(\nu)+a_5^{\log}\log u)u^4 + \nu (a_6^c(\nu)+a_6^{\log}\log u)u^5$.
Here $a_5^c\equiv a_{5}^{\rm eff}=23.5$ and $a_6^c\equiv a_{6}^{\rm
  eff}=(-110.5 + 129(1-4\nu))\sqrt{1 - (1.5\times
  10^{-5})/(\nu-0.26)^2}$. 
The analytically most complete version of $B$ is defined through
$\bar D\equiv 1/(A B)$, with $\bar{D}_{\rm 4PN}=1+\nu\left[\bar{d}_2u^2 +
  \bar{d}_3(\nu) u^3 + (\bar{d}_4^c +\bar{d}_4^{\rm log}\log
  u)u^4\right]$. Here, contrary to~\cite{Damour:2012ky}, which used the
3PN-accurate $\bar{D}$ function, we use the above 4PN-accurate version
with $\bar{d}_4^c=226$~\cite{Barack:2010ny,Barausse:2011dq} and
$\bar{d}_4^{\rm
  log}=592/15$~\cite{Blanchet:2010zd,Barack:2010ny,Barausse:2011dq}.
Hereafter we will refer to this state-of-the-art EOB model
as ``${\rm EOBNR_{\rm 5PNlog}}$''.
\begin{figure}[t]
  \centering
 \includegraphics[width=.47\textwidth]{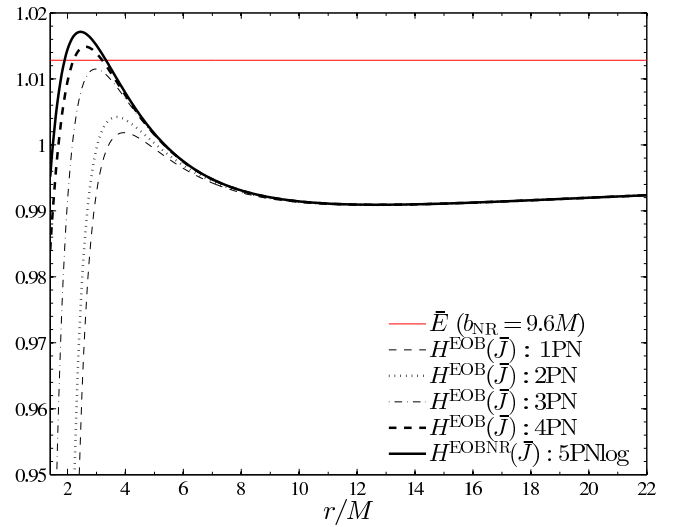} 
    \caption{\label{fig:Ham} EOB effective potentials $H_{\rm EOB}(r,\bar{J},p_r=0)$
      at various PN approximations. The
      values $(\bar{E},\bar{J})=\left(E_{\rm in}-\frac12 \Delta E^{\rm
        NR},J_{\rm in}^{\rm NR}-\frac12 \Delta J^{\rm NR}\right)$
      correspond, in Table~\ref{tab:all_info}, to ${b}_{\rm
        NR}=9.6\,M$. The EOB motion corresponds to a particle starting 
      at large $r$ with negative radial momentum $p_{r}^{0}$ and moving
      towards the left at constant energy $\bar{E}$. Note that 1PN, 2PN 
      and 3PN EOB models predict plunge instead of scattering.}
\end{figure}
We start the integration of the EOB equations of motion 
with $j_0=\bar{J}/(M \mu)$ and $p_{r_*}^0$ obtained by
  solving the equation $\bar{E}=H_{\rm EOB}(r_0,j_0,p_{r_*}^0)$.  
The gauge-invariant scattering angle $\chi^{\rm EOBNR}_{\rm
    5PNlog}(\bar{E},\bar{J})$ obtained from this integration is given in
the fourth column in Table~\ref{tab:EOBNRPN}, to be compared
to the corresponding NR results (third column).
The agreement between these values of $\chi$ is remarkably good. 
The fractional disagreements are equal to: $0.048\%$ for the largest impact
parameter ($b_{\rm NR}=16M$); $1.8\%$ for $b_{\rm NR}=10M$
and  $5.3\%$ for the smallest impact parameter, $b_{\rm NR}=9.6M$ 
(which corresponds to a closest EOB relative distance $r_{\rm min}^{\rm
EOB}=3.3M$)[see Fig.~\ref{fig:angle}].

We (over-)estimate a two-sided uncertainty on
$\chi^{\rm EOBNR}_{\rm 5PNlog}$ \hbox{$\sigma_\chi\equiv
    \pm\frac12 [\langle\chi(E,J)\rangle - \chi(\bar{E},\bar{J})]$} by comparing $\chi(\bar{E},\bar{J})$ to
the average value $\langle \chi \rangle=\frac12(\chi_{\rm in} + \chi_{\rm
  out})$ where 
  $\chi_{\rm in}\equiv \chi(E_{\rm in},J_{\rm
  in})$ and $\chi_{\rm out}\equiv \chi(E_{\rm out},J_{\rm
  out})$. This uncertainty is given in parentheses on the last two
digits in Table~\ref{tab:EOBNRPN}. [In the case ${b}_{\rm NR}=9.6\,M$,
  $E_{\rm out}$ is so small that one cannot use this procedure. In that
  case we estimate an analytical uncertainty from the curvature of the
  function $\chi(E,J)$ around $(\bar{E},\bar{J})$]. Note that, as
  is evident in Fig.~\ref{fig:angle}, if we consider the combined
  uncertainties on $\chi$, the NR and ${\rm EOBNR_{5PNlog}}$ results are
  fully compatible for the entire examined parameter space.

To probe the sensitivity of this NR/EOB agreement on the
precise structure of the EOB Hamiltonian, we analyze the effects of
replacing the 5PN-accurate $A$ and the 4PN-accurate $\bar{D}$ potentials
used in ${\rm EOBNR_{5PNlog}}$ by potentials of lower PN accuracy. We
denote by ${\rm EOB}_{n\rm PN}$ an EOB Hamiltonian defined by
truncating the $A$ and $\bar{D}$ potentials to their $n$PN accuracy. The
result of the corresponding analytical computations of $\chi$ are listed
in Table~\ref{tab:EOBNRPN}. Note that as we lessen the PN accuracy of the
EOB Hamiltonian, the disagreement in $\chi$ increases monotonically (see
Fig.~\ref{fig:angle}). The empty slots in Table~\ref{tab:EOBNRPN}
correspond to configurations where the peak of the EOB effective
potential is lower than $\bar{E}$, so that the analytical evolution
leads to an immediate plunge (see Fig.~\ref{fig:Ham}).

Finally, we explored the sensitivity of the NR/AR comparison on
the resummation procedure built into the EOB formalism, by computing
the predictions for $\chi$ made by the {\it nonresummed} PN-expanded
Hamiltonian. At 3PN accuracy, this (center-of-mass) Hamiltonian (in ADM
coordinates) is a polynomial with 24 terms of the form $(H-M)/\mu \sim
\frac12 p^2-1/r + c^{-2}(p^4 + p^2/r+1/r^2) + c^{-4}(p^8 + \cdots +
1/r^4) + c^{-6}(p^{10}+\cdots + 1/r^4)$
(see~\cite{Jaranowski:1997ky,Damour:2001bu}). The result of computing
$\chi$ (following the same procedure as above) from this 3PN Hamiltonian
is listed in the 9th column of Table~\ref{tab:EOBNRPN}. In addition, the
result of considering 2PN and 1PN truncations of this Hamiltonian is
included in the table. Several conclusions can be drawn from this
comparison. 
First, for most values of the impact  parameter ${b}_{\rm NR}$ the PN$_{\rm 3PN}$/NR
disagreement is significantly larger than the corresponding EOB$_{\rm 3PN}$/NR
disagreement (see Fig.~\ref{fig:angle}). 
For instance, for  ${b}_{\rm NR}=16\,M$, ${\rm PN}_{\rm 3PN}$ 
and NR disagree by $-2.2\%$, while  ${\rm EOB}_{\rm 3PN}$ and NR disagree by $0.20\%$;
and for ${b}_{\rm NR}=10.6\,M$, ${\rm PN}_{\rm 3PN}$ 
and NR disagree by $-31\%$, while  ${\rm EOB}_{\rm 3PN}$ and NR disagree by $+5.9\%$.
Note also the sizable difference between the predictions made
by the 1PN-accurate PN-expanded Hamiltonian and the 1PN-accurate PN-resummed 
EOB Hamiltonian: e.g., for  ${b}_{\rm NR}=16\,M$, the PN prediction disagrees 
with NR by $+54\%$ while the EOB disagreement is just $+3.2\%$.

\paragraph*{Conclusions--}

We have performed the first numerical computation of the scattering angle
$\chi$ of nonspinning, equal-mass BBH encounters varying the impact
parameter while keeping essentially fixed the incoming energy. The range
of explored impact parameter is such that the scattering angle varies
between 70.7 and 305.8 degrees. Correspondingly, the closest distance of
approach of the two BHs (in EOB coordinates) was found to vary between
$10.8\,M$ and $3.3\,M$, indicating that we are indeed exploring the
strong-field dynamics of the BHs. We have compared the NR data to two
different analytical approaches to describing the orbital dynamics of BBHs:
PN theory and the EOB formalism. Our main finding is that, as the impact parameter
$b_{\rm NR}$ decreases, the PN predictions become quite inaccurate 
(by more than a factor 2) while the NR-calibrated EOB predictions keep 
agreeing with NR within their combined error bars. This NR/EOB agreement 
is remarkable since the configurations considered here explore
a dynamical regime, in the $(E,J)$ plane, which is very different from
the quasi-circular configurations used for the calibration of the EOB
model. Note also how the uncalibrated, purely analytical, EOB models
monotonically approach the NR results as their PN accuracy is increased
(see Fig.~\ref{fig:angle}).

Overall, our study opens a new avenue for extracting from NR simulations
nonperturbative information to complete the EOB formalism. In
particular NR scattering experiments for small impact parameters allow
one to probe the height and shape of the EOB effective energy potential
very close to its peak, i.e., for BH separations of the order
of $3M$.

\medskip

\begin{acknowledgments}
  SH, IH, and LR thank Abraham Harte for helpful discussions.
  The computations were performed on the Datura cluster at the AEI and
  on the XSEDE network (allocation TG-MCA02N014). This work was
  supported in part by the DFG grant SFB/Transregio~7
  ``Gravitational-Wave Astronomy''.
\end{acknowledgments}

\bibliographystyle{apsrev4-1-noeprint}
\bibliography{eobnr_scattering}

\begin{thebibliography}{22}%
\makeatletter
\providecommand \@ifxundefined [1]{%
 \@ifx{#1\undefined}
}%
\providecommand \@ifnum [1]{%
 \ifnum #1\expandafter \@firstoftwo
 \else \expandafter \@secondoftwo
 \fi
}%
\providecommand \@ifx [1]{%
 \ifx #1\expandafter \@firstoftwo
 \else \expandafter \@secondoftwo
 \fi
}%
\providecommand \natexlab [1]{#1}%
\providecommand \enquote  [1]{``#1''}%
\providecommand \bibnamefont  [1]{#1}%
\providecommand \bibfnamefont [1]{#1}%
\providecommand \citenamefont [1]{#1}%
\providecommand \href@noop [0]{\@secondoftwo}%
\providecommand \href [0]{\begingroup \@sanitize@url \@href}%
\providecommand \@href[1]{\@@startlink{#1}\@@href}%
\providecommand \@@href[1]{\endgroup#1\@@endlink}%
\providecommand \@sanitize@url [0]{\catcode `\\12\catcode `\$12\catcode
  `\&12\catcode `\#12\catcode `\^12\catcode `\_12\catcode `\%12\relax}%
\providecommand \@@startlink[1]{}%
\providecommand \@@endlink[0]{}%
\providecommand \url  [0]{\begingroup\@sanitize@url \@url }%
\providecommand \@url [1]{\endgroup\@href {#1}{\urlprefix }}%
\providecommand \urlprefix  [0]{URL }%
\providecommand \Eprint [0]{\href }%
\providecommand \doibase [0]{http://dx.doi.org/}%
\providecommand \selectlanguage [0]{\@gobble}%
\providecommand \bibinfo  [0]{\@secondoftwo}%
\providecommand \bibfield  [0]{\@secondoftwo}%
\providecommand \translation [1]{[#1]}%
\providecommand \BibitemOpen [0]{}%
\providecommand \bibitemStop [0]{}%
\providecommand \bibitemNoStop [0]{.\EOS\space}%
\providecommand \EOS [0]{\spacefactor3000\relax}%
\providecommand \BibitemShut  [1]{\csname bibitem#1\endcsname}%
\let\auto@bib@innerbib\@empty
\bibitem [{\citenamefont {Pretorius}\ and\ \citenamefont
  {Khurana}(2007)}]{Pretorius:2007jn}%
  \BibitemOpen
  \bibfield  {author} {\bibinfo {author} {\bibfnamefont {F.}~\bibnamefont
  {Pretorius}}\ and\ \bibinfo {author} {\bibfnamefont {D.}~\bibnamefont
  {Khurana}},\ }\href {\doibase 10.1088/0264-9381/24/12/S07} {\bibfield
  {journal} {\bibinfo  {journal} {Class.Quant.Grav.}\ }\textbf {\bibinfo
  {volume} {24}},\ \bibinfo {pages} {S83} (\bibinfo {year} {2007})}\BibitemShut
  {NoStop}%
\bibitem [{\citenamefont {Shibata}\ \emph {et~al.}(2008)\citenamefont
  {Shibata}, \citenamefont {Okawa},\ and\ \citenamefont
  {Yamamoto}}]{Shibata:2008rq}%
  \BibitemOpen
  \bibfield  {author} {\bibinfo {author} {\bibfnamefont {M.}~\bibnamefont
  {Shibata}}, \bibinfo {author} {\bibfnamefont {H.}~\bibnamefont {Okawa}}, \
  and\ \bibinfo {author} {\bibfnamefont {T.}~\bibnamefont {Yamamoto}},\ }\href
  {\doibase 10.1103/PhysRevD.78.101501} {\bibfield  {journal} {\bibinfo
  {journal} {Phys.Rev.}\ }\textbf {\bibinfo {volume} {D78}},\ \bibinfo {pages}
  {101501} (\bibinfo {year} {2008})}\BibitemShut {NoStop}%
\bibitem [{\citenamefont {Sperhake}\ \emph {et~al.}(2008)\citenamefont
  {Sperhake}, \citenamefont {Cardoso}, \citenamefont {Pretorius}, \citenamefont
  {Berti},\ and\ \citenamefont {Gonzalez}}]{Sperhake:2008ga}%
  \BibitemOpen
  \bibfield  {author} {\bibinfo {author} {\bibfnamefont {U.}~\bibnamefont
  {Sperhake}}, \bibinfo {author} {\bibfnamefont {V.}~\bibnamefont {Cardoso}},
  \bibinfo {author} {\bibfnamefont {F.}~\bibnamefont {Pretorius}}, \bibinfo
  {author} {\bibfnamefont {E.}~\bibnamefont {Berti}}, \ and\ \bibinfo {author}
  {\bibfnamefont {J.~A.}\ \bibnamefont {Gonzalez}},\ }\href {\doibase
  10.1103/PhysRevLett.101.161101} {\bibfield  {journal} {\bibinfo  {journal}
  {Phys.Rev.Lett.}\ }\textbf {\bibinfo {volume} {101}},\ \bibinfo {pages}
  {161101} (\bibinfo {year} {2008})}\BibitemShut {NoStop}%
\bibitem [{\citenamefont {Sperhake}\ \emph {et~al.}(2009)\citenamefont
  {Sperhake}, \citenamefont {Cardoso}, \citenamefont {Pretorius}, \citenamefont
  {Berti}, \citenamefont {Hinderer} \emph {et~al.}}]{Sperhake:2009jz}%
  \BibitemOpen
  \bibfield  {author} {\bibinfo {author} {\bibfnamefont {U.}~\bibnamefont
  {Sperhake}}, \bibinfo {author} {\bibfnamefont {V.}~\bibnamefont {Cardoso}},
  \bibinfo {author} {\bibfnamefont {F.}~\bibnamefont {Pretorius}}, \bibinfo
  {author} {\bibfnamefont {E.}~\bibnamefont {Berti}}, \bibinfo {author}
  {\bibfnamefont {T.}~\bibnamefont {Hinderer}},  \emph {et~al.},\ }\href
  {\doibase 10.1103/PhysRevLett.103.131102} {\bibfield  {journal} {\bibinfo
  {journal} {Phys.Rev.Lett.}\ }\textbf {\bibinfo {volume} {103}},\ \bibinfo
  {pages} {131102} (\bibinfo {year} {2009})}\BibitemShut {NoStop}%
\bibitem [{\citenamefont {Witek}\ \emph {et~al.}(2010)\citenamefont {Witek},
  \citenamefont {Zilhao}, \citenamefont {Gualtieri}, \citenamefont {Cardoso},
  \citenamefont {Herdeiro} \emph {et~al.}}]{Witek:2010xi}%
  \BibitemOpen
  \bibfield  {author} {\bibinfo {author} {\bibfnamefont {H.}~\bibnamefont
  {Witek}}, \bibinfo {author} {\bibfnamefont {M.}~\bibnamefont {Zilhao}},
  \bibinfo {author} {\bibfnamefont {L.}~\bibnamefont {Gualtieri}}, \bibinfo
  {author} {\bibfnamefont {V.}~\bibnamefont {Cardoso}}, \bibinfo {author}
  {\bibfnamefont {C.}~\bibnamefont {Herdeiro}},  \emph {et~al.},\ }\href
  {\doibase 10.1103/PhysRevD.82.104014} {\bibfield  {journal} {\bibinfo
  {journal} {Phys.Rev.}\ }\textbf {\bibinfo {volume} {D82}},\ \bibinfo {pages}
  {104014} (\bibinfo {year} {2010})}\BibitemShut {NoStop}%
\bibitem [{\citenamefont {Sperhake}\ \emph {et~al.}(2013)\citenamefont
  {Sperhake}, \citenamefont {Berti}, \citenamefont {Cardoso},\ and\
  \citenamefont {Pretorius}}]{Sperhake:2012me}%
  \BibitemOpen
  \bibfield  {author} {\bibinfo {author} {\bibfnamefont {U.}~\bibnamefont
  {Sperhake}}, \bibinfo {author} {\bibfnamefont {E.}~\bibnamefont {Berti}},
  \bibinfo {author} {\bibfnamefont {V.}~\bibnamefont {Cardoso}}, \ and\
  \bibinfo {author} {\bibfnamefont {F.}~\bibnamefont {Pretorius}},\ }\href
  {\doibase 10.1103/PhysRevLett.111.041101} {\bibfield  {journal} {\bibinfo
  {journal} {Phys.Rev.Lett.}\ }\textbf {\bibinfo {volume} {111}},\ \bibinfo
  {pages} {041101} (\bibinfo {year} {2013})}\BibitemShut {NoStop}%
\bibitem [{\citenamefont {Buonanno}\ and\ \citenamefont
  {Damour}(1999)}]{Buonanno:1998gg}%
  \BibitemOpen
  \bibfield  {author} {\bibinfo {author} {\bibfnamefont {A.}~\bibnamefont
  {Buonanno}}\ and\ \bibinfo {author} {\bibfnamefont {T.}~\bibnamefont
  {Damour}},\ }\href {\doibase 10.1103/PhysRevD.59.084006} {\bibfield
  {journal} {\bibinfo  {journal} {Phys. Rev.}\ }\textbf {\bibinfo {volume}
  {D59}},\ \bibinfo {pages} {084006} (\bibinfo {year} {1999})}\BibitemShut
  {NoStop}%
\bibitem [{\citenamefont {Buonanno}\ and\ \citenamefont
  {Damour}(2000)}]{Buonanno:2000ef}%
  \BibitemOpen
  \bibfield  {author} {\bibinfo {author} {\bibfnamefont {A.}~\bibnamefont
  {Buonanno}}\ and\ \bibinfo {author} {\bibfnamefont {T.}~\bibnamefont
  {Damour}},\ }\href {\doibase 10.1103/PhysRevD.62.064015} {\bibfield
  {journal} {\bibinfo  {journal} {Phys. Rev.}\ }\textbf {\bibinfo {volume}
  {D62}},\ \bibinfo {pages} {064015} (\bibinfo {year} {2000})}\BibitemShut
  {NoStop}%
\bibitem [{\citenamefont {Damour}\ \emph {et~al.}(2000)\citenamefont {Damour},
  \citenamefont {Jaranowski},\ and\ \citenamefont {Schaefer}}]{Damour:2000we}%
  \BibitemOpen
  \bibfield  {author} {\bibinfo {author} {\bibfnamefont {T.}~\bibnamefont
  {Damour}}, \bibinfo {author} {\bibfnamefont {P.}~\bibnamefont {Jaranowski}},
  \ and\ \bibinfo {author} {\bibfnamefont {G.}~\bibnamefont {Schaefer}},\
  }\href {\doibase 10.1103/PhysRevD.62.084011} {\bibfield  {journal} {\bibinfo
  {journal} {Phys. Rev.}\ }\textbf {\bibinfo {volume} {D62}},\ \bibinfo {pages}
  {084011} (\bibinfo {year} {2000})}\BibitemShut {NoStop}%
\bibitem [{\citenamefont {Loffler}\ \emph {et~al.}(2012)\citenamefont
  {Loffler}, \citenamefont {Faber}, \citenamefont {Bentivegna}, \citenamefont
  {Bode}, \citenamefont {Diener} \emph {et~al.}}]{Loffler:2011ay}%
  \BibitemOpen
  \bibfield  {author} {\bibinfo {author} {\bibfnamefont {F.}~\bibnamefont
  {Loffler}}, \bibinfo {author} {\bibfnamefont {J.}~\bibnamefont {Faber}},
  \bibinfo {author} {\bibfnamefont {E.}~\bibnamefont {Bentivegna}}, \bibinfo
  {author} {\bibfnamefont {T.}~\bibnamefont {Bode}}, \bibinfo {author}
  {\bibfnamefont {P.}~\bibnamefont {Diener}},  \emph {et~al.},\ }\href
  {\doibase 10.1088/0264-9381/29/11/115001} {\bibfield  {journal} {\bibinfo
  {journal} {Class.Quant.Grav.}\ }\textbf {\bibinfo {volume} {29}},\ \bibinfo
  {pages} {115001} (\bibinfo {year} {2012})}\BibitemShut {NoStop}%
\bibitem [{Cactus developers()}]{Cactuscode:web}%
  \BibitemOpen
  Cactus developers,\ \href {http://www.cactuscode.org/} {\enquote {\bibinfo
  {title} {{Cactus Computational Toolkit}},}\ }\BibitemShut {NoStop}%
\bibitem [{\citenamefont {Brown}\ \emph {et~al.}(2009)\citenamefont {Brown},
  \citenamefont {Diener}, \citenamefont {Sarbach}, \citenamefont {Schnetter},\
  and\ \citenamefont {Tiglio}}]{Brown:2008sb}%
  \BibitemOpen
  \bibfield  {author} {\bibinfo {author} {\bibfnamefont {J.~D.}\ \bibnamefont
  {Brown}}, \bibinfo {author} {\bibfnamefont {P.}~\bibnamefont {Diener}},
  \bibinfo {author} {\bibfnamefont {O.}~\bibnamefont {Sarbach}}, \bibinfo
  {author} {\bibfnamefont {E.}~\bibnamefont {Schnetter}}, \ and\ \bibinfo
  {author} {\bibfnamefont {M.}~\bibnamefont {Tiglio}},\ }\href {\doibase
  10.1103/PhysRevD.79.044023} {\bibfield  {journal} {\bibinfo  {journal} {Phys.
  Rev. D}\ }\textbf {\bibinfo {volume} {79}},\ \bibinfo {pages} {044023}
  (\bibinfo {year} {2009})}\BibitemShut {NoStop}%
\bibitem [{\citenamefont {Thornburg}(2004)}]{Thornburg:2003sf}%
  \BibitemOpen
  \bibfield  {author} {\bibinfo {author} {\bibfnamefont {J.}~\bibnamefont
  {Thornburg}},\ }\href {\doibase 10.1088/0264-9381/21/2/026} {\bibfield
  {journal} {\bibinfo  {journal} {Class. Quantum Grav.}\ }\textbf {\bibinfo
  {volume} {21}},\ \bibinfo {pages} {743} (\bibinfo {year} {2004})}\BibitemShut
  {NoStop}%
\bibitem [{\citenamefont {Schnetter}\ \emph {et~al.}(2004)\citenamefont
  {Schnetter}, \citenamefont {Hawley},\ and\ \citenamefont
  {Hawke}}]{Schnetter:2003rb}%
  \BibitemOpen
  \bibfield  {author} {\bibinfo {author} {\bibfnamefont {E.}~\bibnamefont
  {Schnetter}}, \bibinfo {author} {\bibfnamefont {S.~H.}\ \bibnamefont
  {Hawley}}, \ and\ \bibinfo {author} {\bibfnamefont {I.}~\bibnamefont
  {Hawke}},\ }\href {\doibase 10.1088/0264-9381/21/6/014} {\bibfield  {journal}
  {\bibinfo  {journal} {Class. Quantum Grav.}\ }\textbf {\bibinfo {volume}
  {21}},\ \bibinfo {pages} {1465} (\bibinfo {year} {2004})}\BibitemShut
  {NoStop}%
\bibitem [{\citenamefont {Ansorg}\ \emph {et~al.}(2004)\citenamefont {Ansorg},
  \citenamefont {Br{\"u}gmann},\ and\ \citenamefont {Tichy}}]{Ansorg:2004ds}%
  \BibitemOpen
  \bibfield  {author} {\bibinfo {author} {\bibfnamefont {M.}~\bibnamefont
  {Ansorg}}, \bibinfo {author} {\bibfnamefont {B.}~\bibnamefont
  {Br{\"u}gmann}}, \ and\ \bibinfo {author} {\bibfnamefont {W.}~\bibnamefont
  {Tichy}},\ }\href {\doibase 10.1103/PhysRevD.70.064011} {\bibfield  {journal}
  {\bibinfo  {journal} {Phys. Rev. D}\ }\textbf {\bibinfo {volume} {70}},\
  \bibinfo {pages} {064011} (\bibinfo {year} {2004})}\BibitemShut {NoStop}%
\bibitem [{\citenamefont {Bini}\ and\ \citenamefont
  {Damour}(2012)}]{Bini:2012ji}%
  \BibitemOpen
  \bibfield  {author} {\bibinfo {author} {\bibfnamefont {D.}~\bibnamefont
  {Bini}}\ and\ \bibinfo {author} {\bibfnamefont {T.}~\bibnamefont {Damour}},\
  }\href {\doibase 10.1103/PhysRevD.86.124012} {\bibfield  {journal} {\bibinfo
  {journal} {Phys.Rev.}\ }\textbf {\bibinfo {volume} {D86}},\ \bibinfo {pages}
  {124012} (\bibinfo {year} {2012})}\BibitemShut {NoStop}%
\bibitem [{\citenamefont {Damour}\ \emph {et~al.}(2013)\citenamefont {Damour},
  \citenamefont {Nagar},\ and\ \citenamefont {Bernuzzi}}]{Damour:2012ky}%
  \BibitemOpen
  \bibfield  {author} {\bibinfo {author} {\bibfnamefont {T.}~\bibnamefont
  {Damour}}, \bibinfo {author} {\bibfnamefont {A.}~\bibnamefont {Nagar}}, \
  and\ \bibinfo {author} {\bibfnamefont {S.}~\bibnamefont {Bernuzzi}},\ }\href
  {\doibase 10.1103/PhysRevD.87.084035} {\bibfield  {journal} {\bibinfo
  {journal} {Phys.Rev.}\ }\textbf {\bibinfo {volume} {D87}},\ \bibinfo {pages}
  {084035} (\bibinfo {year} {2013})}\BibitemShut {NoStop}%
\bibitem [{\citenamefont {Barack}\ \emph {et~al.}(2010)\citenamefont {Barack},
  \citenamefont {Damour},\ and\ \citenamefont {Sago}}]{Barack:2010ny}%
  \BibitemOpen
  \bibfield  {author} {\bibinfo {author} {\bibfnamefont {L.}~\bibnamefont
  {Barack}}, \bibinfo {author} {\bibfnamefont {T.}~\bibnamefont {Damour}}, \
  and\ \bibinfo {author} {\bibfnamefont {N.}~\bibnamefont {Sago}},\ }\href
  {\doibase 10.1103/PhysRevD.82.084036} {\bibfield  {journal} {\bibinfo
  {journal} {Phys.Rev.}\ }\textbf {\bibinfo {volume} {D82}},\ \bibinfo {pages}
  {084036} (\bibinfo {year} {2010})}\BibitemShut {NoStop}%
\bibitem [{\citenamefont {Barausse}\ \emph {et~al.}(2012)\citenamefont
  {Barausse}, \citenamefont {Buonanno},\ and\ \citenamefont
  {Le~Tiec}}]{Barausse:2011dq}%
  \BibitemOpen
  \bibfield  {author} {\bibinfo {author} {\bibfnamefont {E.}~\bibnamefont
  {Barausse}}, \bibinfo {author} {\bibfnamefont {A.}~\bibnamefont {Buonanno}},
  \ and\ \bibinfo {author} {\bibfnamefont {A.}~\bibnamefont {Le~Tiec}},\ }\href
  {\doibase 10.1103/PhysRevD.85.064010} {\bibfield  {journal} {\bibinfo
  {journal} {Phys.Rev.}\ }\textbf {\bibinfo {volume} {D85}},\ \bibinfo {pages}
  {064010} (\bibinfo {year} {2012})}\BibitemShut {NoStop}%
\bibitem [{\citenamefont {Blanchet}\ \emph {et~al.}(2010)\citenamefont
  {Blanchet}, \citenamefont {Detweiler}, \citenamefont {Le~Tiec},\ and\
  \citenamefont {Whiting}}]{Blanchet:2010zd}%
  \BibitemOpen
  \bibfield  {author} {\bibinfo {author} {\bibfnamefont {L.}~\bibnamefont
  {Blanchet}}, \bibinfo {author} {\bibfnamefont {S.~L.}\ \bibnamefont
  {Detweiler}}, \bibinfo {author} {\bibfnamefont {A.}~\bibnamefont {Le~Tiec}},
  \ and\ \bibinfo {author} {\bibfnamefont {B.~F.}\ \bibnamefont {Whiting}},\
  }\href {\doibase 10.1103/PhysRevD.81.084033} {\bibfield  {journal} {\bibinfo
  {journal} {Phys.Rev.}\ }\textbf {\bibinfo {volume} {D81}},\ \bibinfo {pages}
  {084033} (\bibinfo {year} {2010})}\BibitemShut {NoStop}%
\bibitem [{\citenamefont {Jaranowski}\ and\ \citenamefont
  {Schaefer}(1998)}]{Jaranowski:1997ky}%
  \BibitemOpen
  \bibfield  {author} {\bibinfo {author} {\bibfnamefont {P.}~\bibnamefont
  {Jaranowski}}\ and\ \bibinfo {author} {\bibfnamefont {G.}~\bibnamefont
  {Schaefer}},\ }\href {\doibase 10.1103/PhysRevD.57.7274,
  10.1103/PhysRevD.63.029902} {\bibfield  {journal} {\bibinfo  {journal}
  {Phys.Rev.}\ }\textbf {\bibinfo {volume} {D57}},\ \bibinfo {pages} {7274}
  (\bibinfo {year} {1998})}\BibitemShut {NoStop}%
\bibitem [{\citenamefont {Damour}\ \emph {et~al.}(2001)\citenamefont {Damour},
  \citenamefont {Jaranowski},\ and\ \citenamefont {Schaefer}}]{Damour:2001bu}%
  \BibitemOpen
  \bibfield  {author} {\bibinfo {author} {\bibfnamefont {T.}~\bibnamefont
  {Damour}}, \bibinfo {author} {\bibfnamefont {P.}~\bibnamefont {Jaranowski}},
  \ and\ \bibinfo {author} {\bibfnamefont {G.}~\bibnamefont {Schaefer}},\
  }\href {\doibase 10.1016/S0370-2693(01)00642-6} {\bibfield  {journal}
  {\bibinfo  {journal} {Phys.Lett.}\ }\textbf {\bibinfo {volume} {B513}},\
  \bibinfo {pages} {147} (\bibinfo {year} {2001})}\BibitemShut {NoStop}%
\end{thebibliography}%

\end{document}